\documentstyle[11pt,newpasp,twoside,epsf]{article}
\def\edcomment#1{\iffalse\marginpar{\raggedright\sl#1\/}\else\relax\fi}
\reversemarginpar

\begin{document}
\title{The orbital period distribution of wide binary millisecond pulsars}
\author{B. Willems}
\affil{Northwestern University, Department of Physics and Astronomy,
    2145 Sheridan Road, Evanston, IL 60208, USA}
\author{U. Kolb}
\affil{Department of Physics and Astronomy, The Open University,
    Walton Hall, Milton Keynes, MK7 6AA, UK}

\begin{abstract}
We present results of a binary population synthesis study on the
orbital period distribution of wide binary millisecond pulsars forming
through four evolutionary channels. In three of the
channels, the progenitor of the millisecond pulsar
undergoes a common envelope phase prior to the supernova explosion
which gives birth to the neutron star. In the fourth channel, the
primary avoids the common-envelope phase and forms a neutron star when
it ascends the asymptotic giant branch. The four formation channels
yield an orbital period distribution which typically shows a
short-period peak below 10 days, a long-period peak around 100 days,
and a cut-off near 200 days. The agreement with the orbital period
distribution of observed binary millisecond pulsars in the Galactic
disk is best when the common-envelope ejection is efficient, the
mass-transfer phase responsible for spinning up the pulsar is highly
non-conservative, and no or moderate supernova kicks are imparted to
neutron stars at birth.
\end{abstract}

\section{Introduction}

In close binaries containing a neutron star (NS) and a non-compact
star, the evolution of the latter and/or the orbit may drive the
system into a semi-detached state where the non-compact star transfers
mass to the NS. If the NS is able to accrete some of the transferred
mass, the associated transfer of angular momentum will spin the NS up
until a millisecond pulsar is born. At the end
of the mass-transfer phase, the donor star's core is exposed as a
white dwarf (WD) and the binary appears as a binary millisecond pulsar
(BMSP). If the mass-transfer phase takes place when the donor is
on the giant branch, the relation between the core mass and 
radius of the donor star, and the relation between the Roche-lobe radius
and the orbital separation, lead to a correlation between the orbital
period of the BMSP and the WD mass (e.g. Joss, Rappaport \&
Lewis 1987).

In this paper, we compare the orbital period distribution of simulated
wide BMSPs from the BiSEPS ({\it Bi}nary {\it S}tellar {\it E}volution
and {\it P}opulation {\it S}ynthesis) code with the orbital period
distribution of BMSPs observed in the Galactic disk. We particularly
investigate whether a set of binary evolution parameters can
be found which is able to reproduce the observed distribution without
including observational selection effects or detailed pulsar physics.

\section{Binary population synthesis}

We consider the evolution of a set of ZAMS binaries with initial
component masses between $0.1\,M_\odot$ and $60\,M_\odot$, and initial
orbital periods ranging from 10 to 10\,000 days, up to a maximum
evolutionary age of 15\,Gyr. The initial primary masses are assumed to
be distributed according to a Salpeter type initial mass function
$\propto M_1^{-2.7}$ for $M_1 \ge 0.75\, M_\odot$, the initial mass
ratio distribution is assumed to be flat, and the initial orbital
separation distribution is assumed to be logarithmically flat.

A full description of the BiSEPS binary population synthesis code can
be found in Willems \& Kolb (2002). We here therefore only briefly
recall the treatment of a few key processes in the formation of wide
BMSPS:
\begin{itemize} 
\item Common-envelope (CE) phases are treated in the usual way by
equating the binding energy of the envelope to the orbital energy
released by the in-spiraling component stars. The outcome of the phase
depends on the envelope-ejection efficiency $\alpha_{\rm CE}$ in the
sense that larger values of $\alpha_{\rm CE}$ yield larger post-CE
orbital separations. In our standard model, we set $\alpha_{\rm
CE}=1.0$.
\item Asymmetric supernova (SN) explosions are assumed to impart a
kick velocity to the newborn NS's center-of-mass. The direction of the
kick velocity is assumed to be distributed isotropically in space,
while the magnitude is drawn from a Maxwellian distribution with a
velocity dispersion of $190\, {\rm km/s}$. The post-SN orbital
parameters of the binaries surviving the SN explosion are determined
as in Kalogera (1996).
\item NSs are thought to have a low accretion efficiency; this is
mimicked by imposing an upper limit $\left(\Delta M_{\rm NS}
\right)_{\rm max} = 0.2\,M_\odot$ on the mass that can be accreted by
a NS. As long as this upper limit is not reached, we adopt an average
mass-accretion rate given by $\dot{M}_{\rm NS} = ( \Delta M_{\rm NS}
)_{\rm max} |\dot{M}_{\rm d}|/M_{\rm d}$, where $M_d$ is the mass of
the donor star. In addition, the NS accretion rate is limited to the
Eddington rate at all times.
\end{itemize}

\section{Evolutionary channels}

We consider four evolutionary channels leading to the formation of
wide BMSPs (for a review see, e.g., Bhattacharya \& van den Heuvel
1991). The majority of the systems is found to descend from binaries
in which the primary (the initially most massive star) undergoes a
CE phase prior to the formation of the NS. If the post-SN
binary undergoes a stable mass-transfer phase from a low-mass giant
(channel "He~N") or a thermal-timescale mass-transfer phase from an
intermediate mass star followed by a stable mass-transfer phase from a
low-mass star (channel "He~T"), the system evolves into a BMSP
containing a He WD. If, on the other hand, the binary undergoes a
thermal-timescale mass-transfer phase from an intermediate mass star
which is not followed by a stable mass-transfer phase, the system
evolves into a BMSP containing a C/O WD (channel "CO~T"). In the
fourth formation channel, the BMSPs descend from binaries which do not
form a CE prior to the formation of the NS. These systems
rely on suitably directed kicks at the birth of the NS to decrease the
orbital separation sufficiently for the low-mass companion to fill its
Roche lobe on the AGB (see also Kalogera 1998). The outcome is a wide
BMSP containing a C/O WD (channel "CO~N"). The orbital periods and WD
masses of wide BMSPs evolving through the four formation channels are
shown in Fig.~1.

\begin{figure}
\plotfiddle{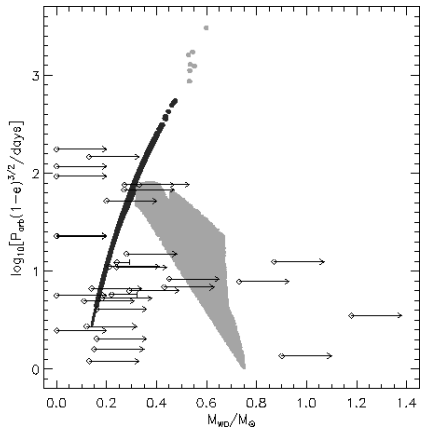}{4.9cm}{0}{35}{35}{-100}{-65}
\caption{Orbital periods vs. WD masses of wide BMSPs. Dark and light
  dots correspond to simulated BMSPs containing He and C/O WDs,
  respectively.  Diamonds and arrows represent the observed Galactic
  BMSPs with the associated error bars.}
\end{figure}

\section{Orbital period distributions}
 
The observed and simulated orbital period distributions of Galactic
BMSPs are shown in Fig.~2. The observed distribution is characterised
by a short- and long-period peak around 10 and 100 days, respectively,
a gap between 30 and 60 days, and a cut-off around 200 days. The
simulated orbital period distributions are generally dominated by
systems forming through the CO~T channel.  However, it is uncertain
whether the accretion process during the thermal time scale
mass-transfer phase giving rise to these systems is efficient enough
to spin the NS up to a MSP. The systems forming through the CO~N
channel, on the other hand, provide the smallest contribution to the
population of wide BMSPs and have periods that are significantly
beyond the 200-day cut-off period in the observed orbital period
distribution. For the remainder of the paper, we therefore focus on
the BMSPs forming through the He~N and He~T channels.

The orbital period distribution of BMSPs forming through channels He~N
and He~T show a peak at long and short orbital periods,
respectively. We tentatively identify these two subpopulations of
BMSPs with the systems comprising the long- and short-period peak in
the observed orbital period distribution. The long-period cut-off
around 200 days is found irrespective of the accretion efficiency of
NSs and results from the upper limit on the initial orbital
periods beyond which the binary avoids the CE phase prior
to the supernova explosion of the primary. Consequently, the cut-off
period tends to increase with increasing values of $\alpha_{\rm
CE}$. The short-period peak, on the other hand, is fairly insensitive
to the values of $\alpha_{\rm CE}$ and $\left(\Delta M_{\rm NS}
\right)_{\rm max}$. The kick-velocity dispersion, finally, determines
the spread in the orbital periods after the SN explosion of the
primary. In particular, small velocity dispersion yield narrow
distributions of post-SN orbital periods, while large velocity
dispersions yield wide distributions with relatively more systems at
short orbital periods (see Fig. 7 in Kalogera 1996). These tendencies
are clearly visible when comparing the orbital period distribution of
our standard model with the orbital period distribution of a model
where no kicks are imparted to NSs at birth (see the left- and
right-hand panels of Fig.~2, respectively).

\begin{figure}
\plottwo{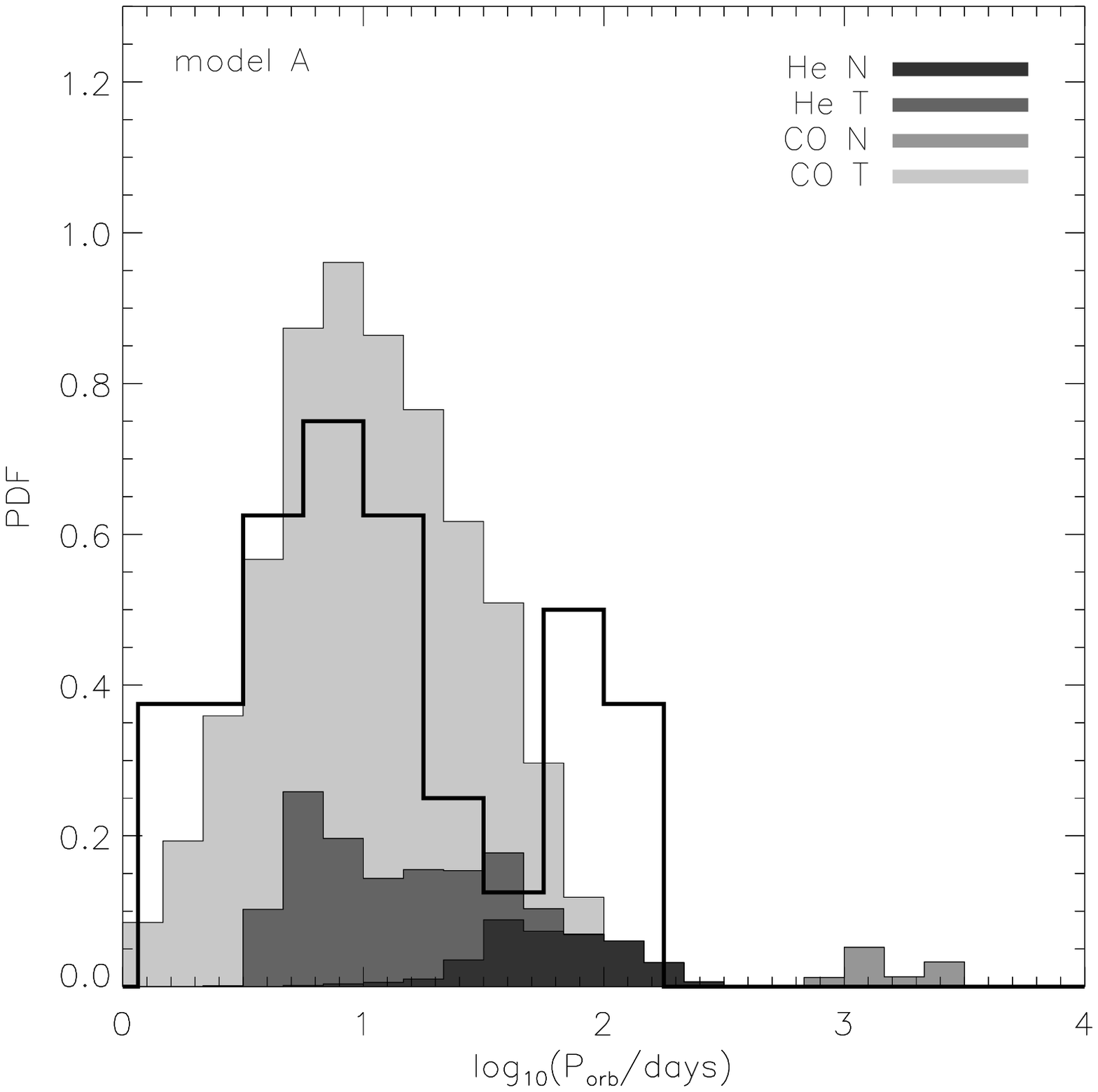}{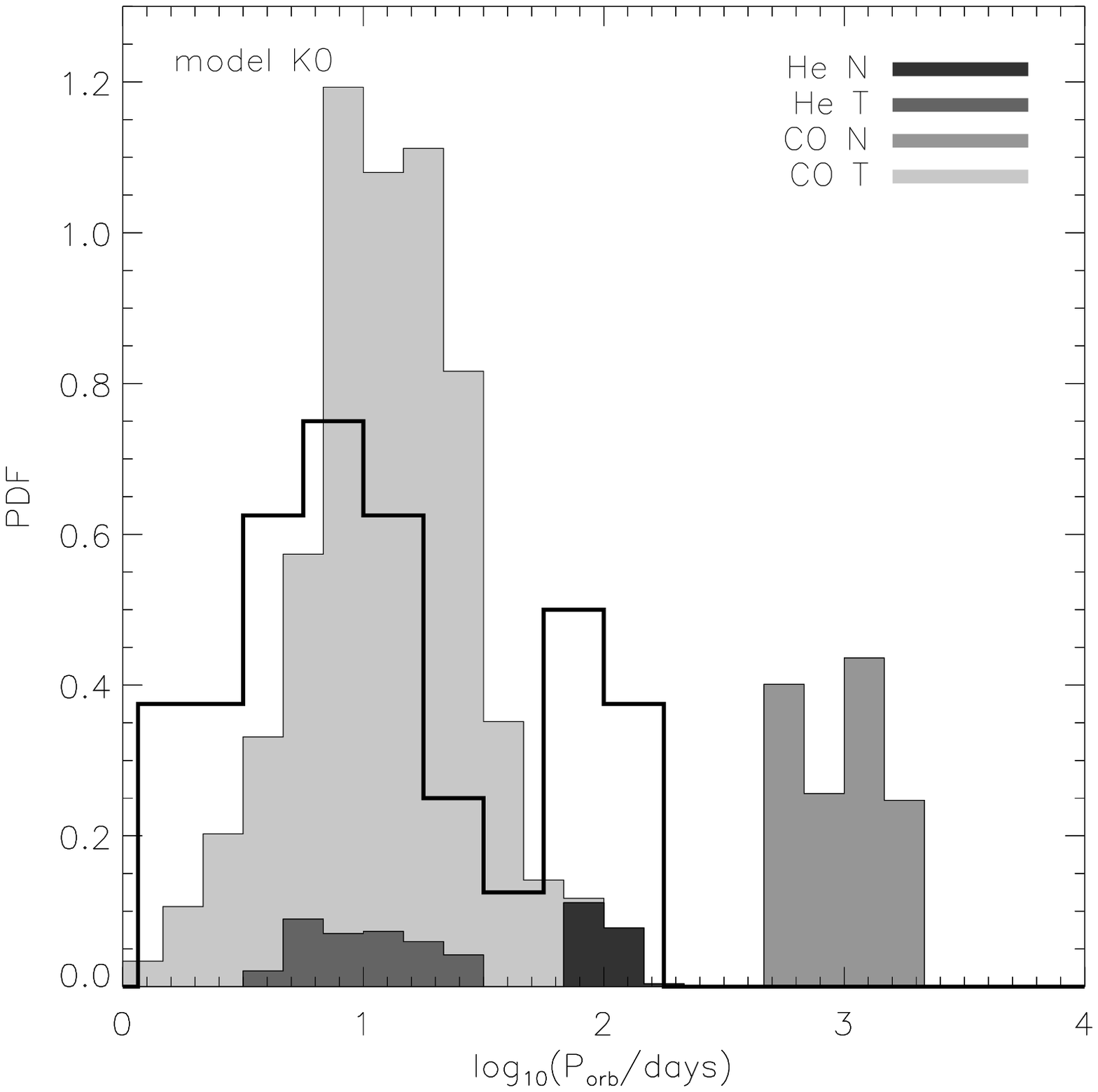}
\caption{Orbital period distribution of observed (solid line) and
  simulated (gray histograms) wide BMSPs. For clarity, the
  distribution functions of BMSPs forming through channels He~N and
  CO~N are multiplied by factors of~5 and~100, respectively. Left
  panel: standard set of parameters; right panel: no kicks imparted to 
  NSs at birth.}
\end{figure}

For conclusion, the observed orbital period distribution is best
reproduced by simulations involving highly non-conservative mass
transfer, common-envelope efficiencies equal to or larger than unity,
and no or moderate supernova kicks at the birth of the NS.
Even though the general effects of varying the population synthesis
model assumptions are fairly clear, there are still too many
uncertainties to come up with a unique set of model parameters giving
the best possible representation of the observed wide BMSP orbital
period distribution (see also Pfahl et al. 2003). As an example of a
"best-fit" solution, we show the orbital period distribution resulting
from a simulation in which we adopted a Maxwellian kick-velocity
distribution with a velocity dispersion of $100\,{\rm km/s}$, a
common-envelope ejection efficiency equal to 1, and an upper limit on
the mass that can be accreted by a NS of $0.2\,M_\odot$. In
addition, we imposed a lower limit of $0.1\,M_\odot$ on the mass that
needs to be accreted to spin the NS up to millisecond
periods. This condition reduces the otherwise dominant channel
CO~T. The initial mass ratio distribution, finally, was taken to be
$\propto 1/q$, which boosts the relative number of systems forming via
mass transfer from low-mass giant stars (the He~N channel). The model
reproduces the observed short- and long-period peaks as well as the
observed period gap, but it fails to produce BMSPs with orbital
periods of 1--3 days. This potential problem may be resolved if the
orbital angular momentum carried away by mass leaving the system
during non-conservative mass transfer is assumed to come from the
donor star instead of the accretor (Nelson, these proceedings), or if
the formation of BMSPs via accretion induced collapse is properly
taken into account (Belczynski \& Taam, in preparation). Our "best-fit"
model also yields no systems containing C/O white dwarfs at orbital
periods shorter than $\sim 30$ days, which may pose a problem to
explain some of the observed systems with massive white dwarfs and
orbital periods shorter than 20 days (cf. Fig. 1). This potential
problem may be resolved by improved modelling of thermal-timescale
mass-transfer phases involving Hertzsprung-gap donor stars.

\begin{figure}
\plotfiddle{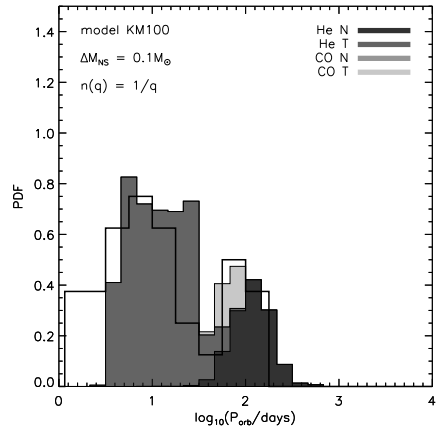}{4.9cm}{0}{35}{35}{-100}{-30}
\caption{Example of a "best-fit" simulated orbital period
distribution of Galactic wide BMSPs (gray histograms).}
\end{figure}

\acknowledgments

BW acknowledges the support of NASA ATP grant NAG5-13236 to Vicky
Kalogera. Part of this research was supported by PPARC.

\end{document}